# A multi-ansatz variational quantum solver for compressible flows

Shaobo Yao[1], Zhiyu Duan[1], Ziteng Wang[1], Wenwen Zhao[1*], and Shiying Xiong[1,2]

[1]*School of Aeronautics and Astronautics, Zhejiang University, Hangzhou 310027, China;*
[2]*National Key Laboratory of Aerospace Physics in Fluids, Mianyang 621000, China*

Simulating nonlinear partial differential equations (PDEs) such as the Navier–Stokes (NS) equations remains computationally intensive, especially when implicit time integration is used to capture multiscale flow dynamics. This work introduces a hybrid quantum–classical framework for solving the linear systems arising from such implicit schemes in compressible flow simulations. At its core is a variational quantum linear solver (VQLS) enhanced by a multi-ansatz tree architecture, designed to expand the accessible solution space and alleviate training issues such as barren plateaus. The proposed method is evaluated through one-dimensional shock tube simulations implemented on a quantum virtual machine. Results demonstrate that the solver accurately captures shock, rarefaction, and contact discontinuities across a range of test cases. Parametric studies further show that increasing the number of ansatz branches and applying domain decomposition improves convergence and stability, even under limited qubit resources. These findings suggest that multi-ansatz VQLS architectures offer a promising pathway for incorporating quantum computing into computational fluid dynamics (CFD), with compatibility for both current noisy intermediate-scale quantum (NISQ) hardware and future fault-tolerant devices.

**Variational quantum algorithms, quantum ansatz tree, Navier–Stokes equations, shock tube simulations**

## 1.   Introduction

Computational fluid dynamics (CFD) provides a numerical framework for solving discretised governing equations of fluid motion and is widely used for analysing and predicting complex flow phenomena. The advancement of CFD has historically relied on progress in high-performance computing (HPC), where increases in classical computational capacity have enabled the resolution of increasingly sophisticated models. However, the continuing miniaturisation of silicon-based processors, as described by Moore's law, is approaching fundamental physical and engineering limits. These constraints pose new challenges to further scaling of classical computational resources and motivate the exploration of alternative computing paradigms. In this context, quantum computing (QC) has emerged as a potentially transformative technology for scientific computing. Its prospective impact on CFD is recognised in NASA's CFD vision 2030 report [1], which identifies QC as a promising direction for addressing the long-term computational demands of fluid simulation.

The conceptual foundations of QC were introduced by Feynman [2], whose vision of simulating physical systems using quantum processors has since inspired the development of numerous quantum algorithms. Among the most notable are Shor's algorithm for integer factorisation [3] and Grover's algorithm for unstructured search [4], both of which demonstrated exponential or quadratic speedups relative to their classical counterparts. As quantum hardware has evolved, Preskill [5] introduced the term "noisy intermediate-scale quantum" (NISQ) to describe the current generation of quantum devices, characterised by limited qubit numbers and significant gate noise. While these devices remain constrained in capability, they offer a practical platform for hybrid quantum–classical algorithms and near-term algorithm development. Givi et al. [6] provide a comprehensive review of QC's implications for aerospace and CFD, advocating long-term engagement with this evolving technology rather than viewing it as a short-term trend.

Recent research has examined the potential application of quantum algorithms to problems in CFD. Surveys by Gaitan [7], Succi [8], and Bharadwaj [9] provide general overviews of developments in this area. Various heuristic and hybrid quantum algorithms have been proposed for fluid-related computations [10-21]. Among them, the Harrow–Hassidim–Lloyd (HHL) algorithm [22] has received attention due to its capacity to solve systems of linear equations, which are frequently encountered in CFD solvers, with potential computational benefits under certain conditions [23].

Other studies have explored quantum-specific formulations of fluid problems. Meng [24] presented an approach based on the hydrodynamic Schrödinger equation, while Liu [25] implemented a convection–diffusion solver on quantum hardware using a classical FTCS discretisation. Todorova et al. [26] applied quantum algorithms originally developed for the Dirac equation to the collisionless Boltzmann equation.

Variational quantum algorithms (VQAs) [27] have been investigated in the context of CFD due to their compatibility with near-term intermediate-scale quantum (NISQ) hardware. These algorithms rely on parameterized quantum circuits combined with classical optimization, and have been applied in several CFD-related studies [28-34]. Meng et al. [35] reported simulations of two-dimensional compressible and decaying vortex flows using Hamiltonian-based techniques on superconducting quantum devices, suggesting that such approaches may be used for more complex flow scenarios.

Quantum machine learning (QML) [36], which combines methods from QC and machine learning has also been considered for fluid simulation [37]. Some QML frameworks have been applied to the solution of nonlinear [38] and partial differential equations (PDEs) [39] arising in CFD, with particular focus on scaling properties. Alternative formulations that are more amenable to QC have also been proposed. For example, quantum Monte Carlo techniques have been shown to provide quadratic speedups in some contexts [40], and possible applications to turbulent mixing have been explored [41].

In addition to stand-alone quantum solvers, several efforts have focused on integrating quantum algorithms within classical CFD frameworks. Meng et al. [42] outlined three central challenges in developing quantum CFD algorithms: (1) the encoding of flow states into quantum registers, (2) the simulation of nonlinear and non-unitary evolution, and (3) the efficient extraction of flow statistics from quantum measurements. The lattice Boltzmann method (LBM), due to its decoupling of streaming and collision processes, has attracted interest as a quantum-compatible approach. Ljubomir [43] implemented a stream function–vorticity formulation based on LBM in Qiskit, demonstrating its performance on two-dimensional cavity flow. Mezzacapo et al. [44] proposed a quantum simulator based on pseudospin–boson systems to encode transport phenomena in a lattice kinetic framework. More recently, Wang et al. [45] introduced a quantum lattice Boltzmann method (QLBM) that combines the dimensional efficiency of LBM with the linear structure of lattice gas cellular automata (LGCA), enabling quantum processing of fluid transport in moderate-dimensional systems.

Despite these developments, several limitations persist. Many quantum solvers remain restricted to linear systems and do not address the nonlinear, multiscale character of the Navier–Stokes (NS) equations. Furthermore, the choice of quantum circuit architecture, or ansatz, plays a critical role in determining algorithm performance [46]. Common layered ansatz may lack sufficient expressivity and often encounter optimisation challenges such as barren plateaus. In addition, most existing studies do not incorporate implicit time integration schemes, which are widely used in CFD and involve the solution of time-dependent linear systems at each timestep.

To address these gaps, the present work develops a variational quantum linear solver (VQLS) framework tailored to implicit discretisation and compressible flow simulation. The method incorporates a multi-ansatz tree structure, designed to enhance representational flexibility and training stability, and is compatible with current NISQ device constraints such as circuit depth and gate fidelity. The framework is validated on a canonical one-dimensional shock tube problem, which serves as a benchmark for compressible flow simulation and quantum solver performance.

Section 2 presents the formulation of the VQLS, including its mathematical foundation, the construction of the multi-ansatz tree, and implementation protocols suitable for NISQ devices. The formulation is further extended to represent the implicit discretisation of the nonlinear NS equations. Section 3 provides numerical validation through three principal investigations: (1) simulation of one-dimensional shock tube flow on a quantum virtual machine, (2) parameter studies examining the effects of ansatz tree configuration on numerical accuracy, and (3) analysis of scalability and current hardware constraints. Concluding remarks are summarised in section 4.

## 2. Computational methods

### 2.1 VQLS

VQAs are hybrid quantum–classical methods developed to operate within the constraints of NISQ devices. These algorithms rely on a parameterised quantum circuit, commonly referred to as a variational ansatz, to prepare a trial quantum state that approximates the desired solution, such as the ground state of a Hamiltonian. The parameters of the circuit are updated iteratively through classical optimisation routines. The optimisation seeks to minimise a cost function, typically defined as the expectation value of a problem-dependent observable. This variational framework enables efficient approximations of complex quantum systems while maintaining robustness against noise and decoherence inherent in current hardware.

VQAs are widely employed in quantum chemistry, where they are used to approximate ground-state energies and wave-functions of molecular Hamiltonians. More generally, VQAs provide a framework for approximating eigenstates of Hermitian operators by preparing a parameterised quantum state and optimising its parameters to minimise the expected energy. This framework extends naturally to the solution of linear systems $\mathcal{A}x = b$, by defining a problem-specific Hermitian operator whose ground state encodes the desired solution.

A common construction introduces the Hamiltonian

$$H = \mathcal{A}^\dagger (I - |b\rangle\langle b|)\mathcal{A}, \tag{1}$$

where $|b\rangle$ is the normalised quantum state corresponding to the vector $b$. By construction, any state $|x\rangle$ satisfying $\mathcal{A}|x\rangle = |b\rangle$ minimises the expectation value $\langle x|H|x\rangle$, yielding zero. In particular, the ground state of $H$ satisfies the eigenvalue equation $H|x\rangle = E_0|x\rangle$ with $E_0 = 0$ and thus encodes the solution $x$ up to normalisation.

As illustrated in figure 1, the variational approach begins by choosing a parameterised quantum state $|\psi(\boldsymbol{\theta})\rangle$ within a prescribed ansatz subspace to approximate the ground state of $H$. The parameters $\boldsymbol{\theta}$ are adjusted to minimise the cost function

$$C(\boldsymbol{\theta}) = \langle\psi(\boldsymbol{\theta})|H|\psi(\boldsymbol{\theta})\rangle, \tag{2}$$

which corresponds to the squared residual $\|\mathcal{A}|\psi(\boldsymbol{\theta})\rangle - |b\rangle\|^2$. Upon convergence, the optimised state $|\psi(\boldsymbol{\theta}^*)\rangle$ serves as an approximation to $|x\rangle$, thereby enabling a variational quantum procedure for solving linear systems.

The method adopts a hybrid quantum–classical optimisation framework, wherein the quantum processor evaluates the expectation value of a cost function and the classical processor updates the variational parameters accordingly. The performance of this approach depends critically on the expressivity of the variational ansatz and the capacity of the problem-specific Hamiltonian to encode the desired solution within its ground state.

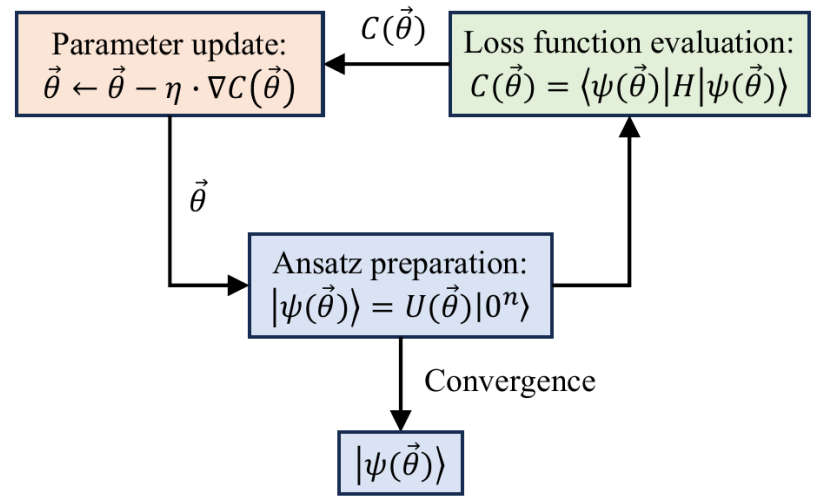

**Figure 1** Quantum ansatz for VQA. Initialize the parameterized quantum state $|\psi(\boldsymbol{\theta})\rangle$, and update the parameter $\boldsymbol{\theta}$ to make it approach the ground state of $H$

VQAs provide a scalable framework for solving high-dimensional problems by enabling efficient exploration of low-dimensional parameter spaces. Their flexibility arises from the freedom to construct Hamiltonians that are tailored to specific applications, rendering the method broadly applicable across a range of domains. Furthermore, the shallow circuit depths and inherent resilience to noise make VQAs particularly well suited for implementation on near-term quantum hardware. Despite these advantages, conventional VQAs suffer from fundamental limitations. In particular, the barren plateau phenomenon, characterised by exponentially vanishing gradients as the system size increases, severely limits the effectiveness of gradient-based optimisation strategies. In addition, the nonconvex structure of the variational landscape frequently leads to convergence towards local rather than global minima, further impeding solution accuracy.

To address these challenges, we introduce a multi-ansatz tree architecture for the solution of quantum linear systems. This formulation employs a dual-level optimisation scheme, in which both the internal variational parameters of each quantum circuit and the external classical coefficients governing their superposition are simultaneously trained. The resulting solution state is expressed as

$$|x\rangle = \sum_{i=1}^{N} \alpha_i |\psi(\boldsymbol{\theta}_i)\rangle, \tag{3}$$

where $\{|\psi(\boldsymbol{\theta}_i)\rangle\}$ denote the outputs of distinct ansatz circuits and $\{\alpha_i\}$ are real coefficients optimised in the classical loop.

As shown in figure 2, each ansatz circuit is constructed from a distinct unitary operator, such as $U_1(\boldsymbol{\theta}_1)$ and $U_2(\boldsymbol{\theta}_2)$, leading to structurally disjoint solution subspaces. This architectural design offers several benefits. It increases the representational capacity of the variational state without incurring additional quantum circuit depth. It enables synergistic opti-misation between quantum and classical components. Most notably, it mitigates barren plateau effects by promoting parameter diversity across the ansatz ensemble.

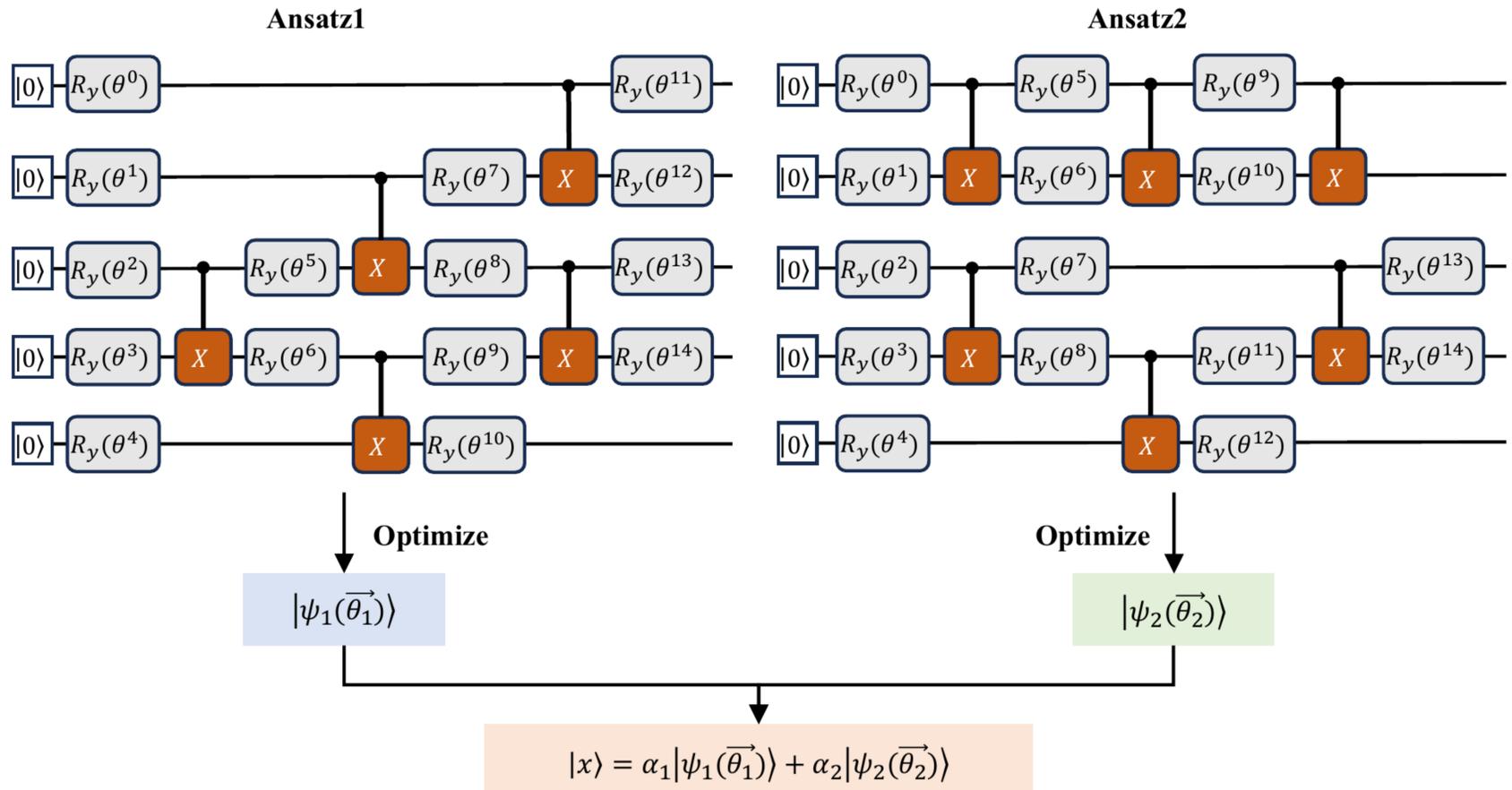

**Figure 2** Schematic representation of the computational procedure for the ansatz tree method. The solution of the equation is decomposed into two independent solution subspaces, each constructed by distinct unitary operators through ansatz circuits.

## 2.2 Discretization of the NS equation

In the continuum description of compressible fluid flow, the NS equations constitute the governing system expressing the conservation of mass, momentum and total energy. For one-dimensional, viscous, compressible flow, the equations may be written in conservative form as

$$\frac{\partial \boldsymbol{Q}}{\partial t} + \frac{\partial \boldsymbol{E}}{\partial x} = \frac{\partial \boldsymbol{E}_v}{\partial x}, \tag{4}$$

where $\boldsymbol{Q}$ denotes the vector of conserved variables, $\boldsymbol{E}$ the inviscid flux vector, and $\boldsymbol{E}_v$ the viscous flux vector, given respectively by

$$\boldsymbol{Q} = \begin{bmatrix} \rho \\ \rho u \\ \rho e \end{bmatrix}, \quad \boldsymbol{E} = \begin{bmatrix} \rho u \\ \rho u^2 + p \\ u(\rho e + p) \end{bmatrix}, \quad \boldsymbol{E}_v = \begin{bmatrix} 0 \\ \tau \\ \kappa \, \partial_x T + u\tau \end{bmatrix}.$$

Here, $\rho$, $u$, $p$, $T$, and $e$ denote the fluid density, velocity, pressure, temperature and specific total energy, respectively. The terms $\tau$ and $\kappa$ represent the viscous stress and thermal conductivity.

To render the governing equations dimensionless, characteristic reference scales are introduced: $\cdots$a length scale $L_0$,

a reference speed of sound $a_\infty$, a reference temperature $T_\infty$, and a reference density $\rho_\infty$. The corresponding nondimensional variables are defined as

$$\begin{cases} \bar{x} = \dfrac{x}{L_0}, \bar{u} = \dfrac{u}{a_\infty}, \bar{p} = \dfrac{p}{\rho_\infty \, a_\infty^2}, \bar{\rho} = \dfrac{\rho}{\rho_\infty}, \\[2ex] \bar{T} = \dfrac{T}{T_\infty}, \bar{e} = \dfrac{e}{a_\infty^2}, \bar{t} = \dfrac{t}{L_0/a_\infty}, \end{cases} \tag{5}$$

where an overbar indicates a nondimensional quantity.

By substituting the nondimensional variables defined in equation (5) into the dimensional form (4), the NS equations are recast in nondimensional form as

$$\frac{\partial \boldsymbol{Q}}{\partial t} + \frac{\partial \boldsymbol{E}}{\partial x} = \frac{Ma_\infty}{Re_\infty} \frac{\partial \boldsymbol{E}_v}{\partial x}, \tag{6}$$

where $Ma_\infty = u_\infty/a_\infty$ and $Re_\infty = \rho_\infty a_\infty L_0/\mu_\infty$ are the freestream Mach and Reynolds numbers, respectively. For clarity, all variables in equation (6) are nondimensional, and the overbars have been omitted hereafter.

The discrete form of the governing equations is expressed as

$$\frac{\Delta \boldsymbol{Q}_i}{\Delta t} = \frac{Ma_\infty}{Re_\infty} \left( \boldsymbol{E}_{v, i+\frac{1}{2}}^{n+1} - \boldsymbol{E}_{v, i-\frac{1}{2}}^{n+1} \right) - \left( \boldsymbol{E}_{i+\frac{1}{2}}^{n+1} - \boldsymbol{E}_{i-\frac{1}{2}}^{n+1} \right), \tag{7}$$

where $\Delta \boldsymbol{Q}_i = \boldsymbol{Q}_i^{n+1} - \boldsymbol{Q}_i^n$, and the superscript $n$ and $n+1$ denote the time iteration steps. The residual, or right-hand side (RHS) term, is defined as

$$\text{RHS}_i^{n+1} = \frac{Ma_\infty}{Re_\infty} \left( \boldsymbol{E}_{v,\,i+\frac{1}{2}}^{n+1} - \boldsymbol{E}_{v,\,i-\frac{1}{2}}^{n+1} \right) - \left( \boldsymbol{E}_{i+\frac{1}{2}}^{n+1} - \boldsymbol{E}_{i-\frac{1}{2}}^{n+1} \right).$$

(8)

The Jacobian matrices of the inviscid and viscous fluxes are defined as

$$\boldsymbol{A} = \frac{\partial \boldsymbol{E}}{\partial \boldsymbol{Q}}, \qquad \boldsymbol{A}_v = \frac{\partial \boldsymbol{E}_v}{\partial \boldsymbol{Q}},$$

(9)

where both are functions of the conserved variables $\boldsymbol{Q}$. Application of a first-order flux vector splitting scheme allows the linearization of the inviscid flux at cell interfaces:

$$\begin{aligned} \boldsymbol{E}_{i+\frac{1}{2}}^{n+1} &= \boldsymbol{E}_{i+\frac{1}{2}}^n + (\boldsymbol{A}\,\Delta\boldsymbol{Q})_{i+\frac{1}{2}} \\ &= \boldsymbol{E}_{i+\frac{1}{2}}^n + \boldsymbol{A}_i^\pm\,\Delta\boldsymbol{Q}_i + \boldsymbol{A}_{i\pm1}^\mp\,\Delta\boldsymbol{Q}_{i\pm1}, \end{aligned}$$

(10)

$$\boldsymbol{E}_{v,\,i+\frac{1}{2}}^{n+1} = \boldsymbol{E}_{v,\,i+\frac{1}{2}}^n + (\boldsymbol{A}_v\,\Delta\boldsymbol{Q})_{i+\frac{1}{2}}.$$

Here, $\boldsymbol{A}_i^\pm = \boldsymbol{R}_i\,\Lambda_i^\pm\,\boldsymbol{L}_i$, where $\boldsymbol{R}_i$ and $\boldsymbol{L}_i = \boldsymbol{R}_i^{-1}$ are the right/left eigenvector matrices of $\boldsymbol{A}_i$, and $\Lambda_i^\pm = \text{diag}\big( \max(\pm\lambda_j, 0) \big)$ is the diagonal matrix of split eigenvalues.

Substituting these linearized expressions into the discrete form results in the following algebraic equation for the implicit time integration:

$$\begin{aligned} \frac{\Delta\boldsymbol{Q}_i}{\Delta t} &+ \left( \boldsymbol{A}_i^+\,\Delta\boldsymbol{Q}_i + \boldsymbol{A}_{i+1}^-\,\Delta\boldsymbol{Q}_{i+1} \right) \\ &- \left( \boldsymbol{A}_{i-1}^+\,\Delta\boldsymbol{Q}_{i-1} + \boldsymbol{A}_i^-\,\Delta\boldsymbol{Q}_i \right) \\ &- \frac{Ma_\infty}{Re_\infty} \left[ (\boldsymbol{A}_v\,\Delta\boldsymbol{Q})_{i+\frac{1}{2}} - (\boldsymbol{A}_v\,\Delta\boldsymbol{Q})_{i-\frac{1}{2}} \right] = \text{RHS}^n. \end{aligned}$$

(11)

The viscous Jacobian is approximated using a central differencing scheme:

$$\begin{aligned} (\boldsymbol{A}_v\,\Delta\boldsymbol{Q})_{i+\frac{1}{2}} &= (\boldsymbol{A}_v\,\Delta\boldsymbol{Q})_{i+1} - (\boldsymbol{A}_v\,\Delta\boldsymbol{Q})_i, \\ (\boldsymbol{A}_v\,\Delta\boldsymbol{Q})_{i-\frac{1}{2}} &= (\boldsymbol{A}_v\,\Delta\boldsymbol{Q})_i - (\boldsymbol{A}_v\,\Delta\boldsymbol{Q})_{i-1}. \end{aligned}$$

(12)

By grouping the terms in equation (11), the resulting linear system may be written compactly as

$$\left( \boldsymbol{D} + \overline{\boldsymbol{U}} + \overline{\boldsymbol{L}} \right) \Delta\boldsymbol{Q} = \text{RHS}^n,$$

(13)

where the diagonal, upper, and lower coefficient matrices are defined as

$$\begin{cases} \boldsymbol{D} = \dfrac{\boldsymbol{I}}{\Delta t} + \varrho(\boldsymbol{A})\,\boldsymbol{I} + 2\dfrac{Ma_\infty}{Re_\infty}\boldsymbol{A}_{v,i}, \\[2mm] \overline{\boldsymbol{U}} = \boldsymbol{A}_{i+1}^- - \dfrac{Ma_\infty}{Re_\infty}\boldsymbol{A}_{v,i+1}, \\[2mm] \overline{\boldsymbol{L}} = -\boldsymbol{A}_{i-1}^+ - \dfrac{Ma_\infty}{Re_\infty}\boldsymbol{A}_{v,i-1}. \end{cases}$$

Here, $\boldsymbol{I}$ is the $3 \times 3$ identity matrix and $\varrho(\boldsymbol{A}) = \boldsymbol{A}_i^+ - \boldsymbol{A}_i^-$ represents the spectral radius of the inviscid Jacobian.

Equation (13) constitutes the fully linearized algebraic system. In conventional CFD solvers, this system is typically solved using the Lower‑Upper Symmetric Gauss‑Seidel (LU-SGS) method, which often represents the dominant computational cost in large-scale simulations. The present study aims to enhance this step through quantum-accelerated linear solvers, enabling a hybrid quantum‑classical framework that integrates quantum routines within standard CFD solution procedures.

### 2.3 VQA for the NS equation

This section presents a VQA designed to approximate the solution of the algebraic system $\mathcal{A}\boldsymbol{x} = \boldsymbol{b}$, which arises from the discretization of the compressible NS equations. The approach is based on variationally preparing a quantum state that encodes the solution, followed by energy minimization to optimize the variational parameters.

The solution vector $\boldsymbol{x}$ is expressed in terms of a complete basis $\{\boldsymbol{e}_i\}$, as $\boldsymbol{x} = \sum_{i=1}^N a_i\boldsymbol{e}_i$, where the coefficients $a_i$ are treated as variational parameters and $N$ is the subspace dimension. Substituting this form into the linear system yields the relation $\sum_{i=1}^N a_i(\mathcal{A}\boldsymbol{e}_i) = \boldsymbol{b}$, which constitutes a reduced-order representation of the original system.

To embed this problem in a quantum framework, a cost functional is defined as a quantum energy expectation:

$$\begin{aligned} E &= \big\| \mathcal{A}|x\rangle - |b\rangle \big\|^2 \\ &= \langle x|\mathcal{A}^\dagger\mathcal{A}|x\rangle - 2\,\text{Re}\,\{\langle b|\mathcal{A}|x\rangle\} + 1. \end{aligned}$$

(14)

Here, $\text{Re}\{\cdot\}$ denotes the real part of a complex quantity.

This formulation enables identification of the optimal quantum state by minimizing the cost functional $E$. The variational process is mathematically equivalent to searching for the ground state of a quantum Hamiltonian.

Recasting equation (14) in light of the constraint $\mathcal{A}\boldsymbol{x} = \boldsymbol{b}$, the energy functional admits an equivalent formulation as

$$E = \langle x,\, H(1)\,x \rangle,$$

(15)

where the Hermitian operator $H(1) = \mathcal{A}^\dagger(I - |b\rangle\langle b|)\mathcal{A}$ serves as an effective Hamiltonian encoding the residual minimization.

The quantum system assumes two key conditions: first, the source vector $|b\rangle$ is prepared by a unitary operator $U_b$ such that $U_b|0^n\rangle = |b\rangle$; second, the matrix $\mathcal{A}$ is decomposed as $\mathcal{A} = \sum_k \beta_k U_k$, with $\beta_k$ denoting real coefficients and $U_k$ unitary operators.

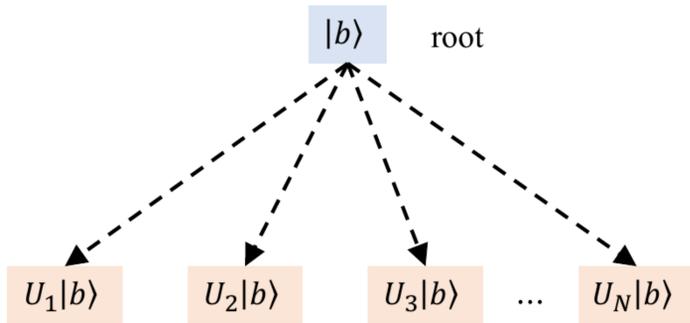

**Figure 3** Structure of the Krylov-based ansatz tree. Each ansatz tree structure corresponds to a single Krylov subspace expansion.

To solve the optimization problem, a parameterized quantum circuit $U_{\text{ansatz}}(\boldsymbol{\theta})$ is employed to generate the trial quantum state:

$$|\psi(\boldsymbol{\theta})\rangle = U_{\text{ansatz}}(\boldsymbol{\theta})|0^n\rangle.$$

The variational optimization proceeds by preparing the quantum state, estimating the expectation value of the cost Hamiltonian, and updating the parameters $\boldsymbol{\theta}$ using classical optimization methods. Convergence is achieved when the cost function reaches a predefined threshold.

To improve solution accuracy and expressivity, a Krylov subspace-enhanced approach is adopted. The solution state is constructed from the span of $\{\boldsymbol{b}, \mathcal{A}\boldsymbol{b}, \mathcal{A}^2\boldsymbol{b}, \ldots, \mathcal{A}^{r-1}\boldsymbol{b}\}$, where $r$ denotes the subspace order. The variational ansatz is structured as a tree, as illustrated in figure 3, where each layer corresponds to a Krylov basis expansion. To control complexity, only subspaces contributing significant gradient information are retained, while others are pruned based on sensitivity analysis. This adaptive mechanism ensures efficient use of quantum resources.

The full VQA framework consists of three components: the parameterized ansatz circuit, a cost function quantifying the residual, and a classical optimizer. The cost function is evaluated via projective measurements as

$$L(\{\alpha_i\}, \{\boldsymbol{\theta}_l\}) = 1 - \frac{\langle x(\{\alpha_i\}, \{\boldsymbol{\theta}_l\})|W|x(\{\alpha_i\}, \{\boldsymbol{\theta}_l\})\rangle}{\langle x(\{\alpha_i\}, \{\boldsymbol{\theta}_l\})|V|x(\{\alpha_i\}, \{\boldsymbol{\theta}_l\})\rangle}, \tag{16}$$

with $W = \mathcal{A}^\dagger|b\rangle\langle b|\mathcal{A}$ and $V = \mathcal{A}^\dagger\mathcal{A}$.

A dual-branch architecture is employed to enhance the ansatz expressivity. Two circuits $U_1(\boldsymbol{\theta}_1)$ and $U_2(\boldsymbol{\theta}_2)$ are initialized with random parameters. The solution state is constructed as a linear combination:

$$|x\rangle = \alpha_1 U_1(\boldsymbol{\theta}_1)|0^n\rangle + \alpha_2 U_2(\boldsymbol{\theta}_2)|0^n\rangle.$$

After convergence, the quantum solution is projected to a classical vector using the normalization

$$x = \epsilon \cdot \frac{|x\rangle}{\||\mathcal{A}|x\rangle\|_2}, \quad \epsilon = \text{sign}\big(\langle x|\mathcal{A}^\dagger|b\rangle\big).$$

The proposed VQA framework offers a scalable pathway to solving fluid dynamics equations using quantum resources. By combining Krylov-based representations with adaptive ansatz growth, the method balances solution accuracy and computational efficiency.

**Table 1** Initial conditions for the shock-tube test cases. Subscripts $L$ and $R$ denote the left and right states, respectively.

| Quantity | Case 1 | Case 2 | Case 3 |
|---|---|---|---|
| $Re_L$ | 0 | 0 | $6.9 \times 10^6$ |
| $Re_R$ | 0 | 0 | $2.2 \times 10^6$ |
| $\rho_L$ | 1.0 | 1.0 | 5.99924 |
| $u_L$ | 0.0 | 0.0 | 19.5975 |
| $p_L$ | 1.0 | 100.0 | 460.894 |
| $\rho_R$ | 0.125 | 1.0 | 5.99242 |
| $u_R$ | 0.0 | 0.0 | −6.19633 |
| $p_R$ | 0.1 | 0.01 | 46.0950 |

## 3. Results and discussions

### 3.1 Numerical setup and experimental procedure

The one-dimensional NS equations are solved under three representative initial configurations to investigate canonical shock-tube dynamics. The selected cases are designed to capture shock wave formation, rarefaction wave propagation, and contact discontinuity evolution, with specific initial conditions listed in table 1. The initial states are defined through a piecewise-constant discontinuity, and viscous effects are included where applicable to assess their influence relative to inviscid Riemann solutions.

For the classical baseline, the resulting linear system $\mathcal{A}\boldsymbol{x} = \boldsymbol{b}$ is solved using LU decomposition. In the VQA, the Adam optimiser is employed with an initial learning rate of $10^{-1}$, a maximum of 600 iterations, and a convergence tolerance of $\delta = 10^{-3}$ for Case 1, and $\delta = 10^{-4}$ for Cases 2 and 3. All simulations are performed on a uniform grid comprising 85 spatial cells, selected to align with the maximum system dimensionality compatible with an 8-qubit quantum register.

To accommodate domain decomposition within the quantum-assisted solver, the computational domain is partitioned into two subregions. The number of ansatz trees is set to two for Case 1 and four for Cases 2 and 3. This configuration enables consistent comparison between classical and hybrid quantum–classical approaches under equivalent spatial resolution and problem dimensionality constraints.

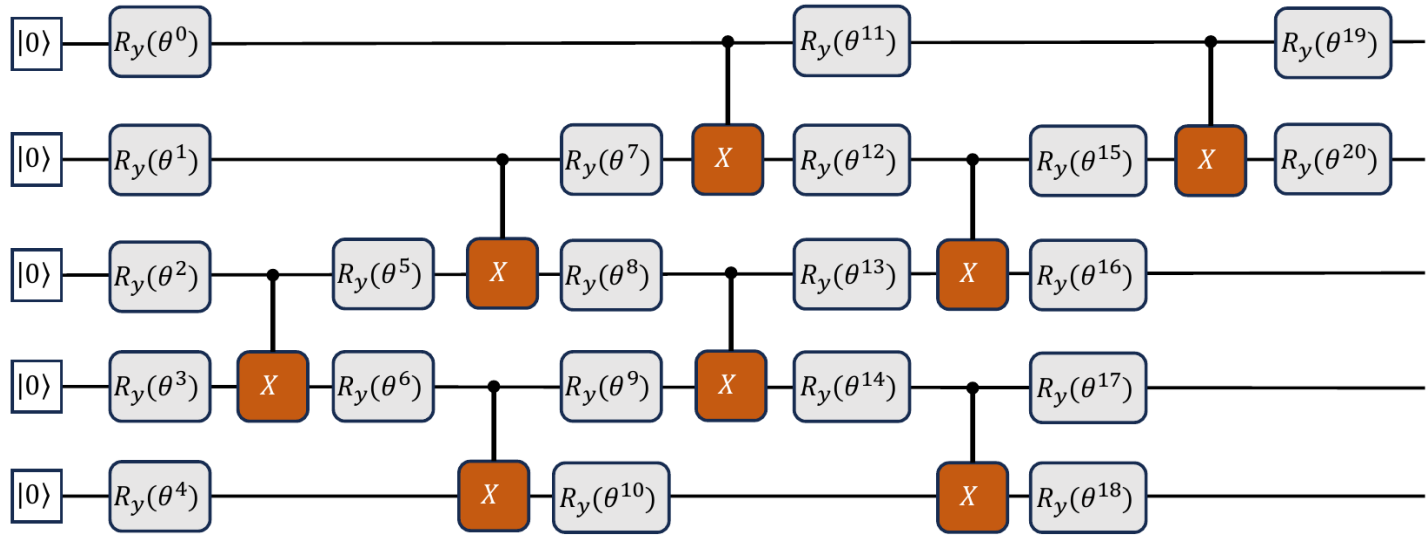

**Figure 4**  Quantum circuit structure for the VQLS-based shock-tube simulation.

### 3.2 Quantum simulation of shock-tube flows

A VQLS, integrated with a multi-ansatz-tree framework, is applied to the one-dimensional shock tube problem. The aim is to recover the spatial distributions of the principal hydrodynamic quantities, namely the density $\rho$, velocity $u$, pressure $p$, and total energy $e$. The quantum circuit configuration used in the simulation is presented in figure 4.

Figures 5–7 present the simulation results for three representative initial configurations. In figure 5, corresponding to case 1, the quantum solver reproduces the classical solution with high fidelity. The density distribution resolves both rarefaction and compression waves, while the velocity and pressure fields closely follow expected discontinuities. The total energy profile remains smooth, preserving consistency with conservation laws.

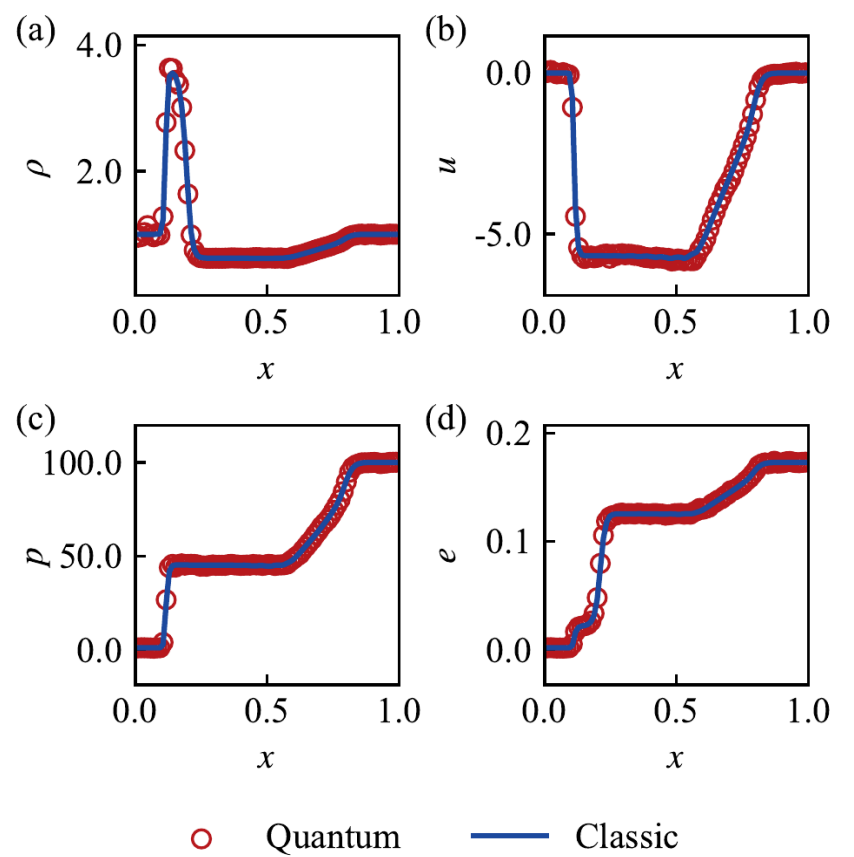

**Figure 6**  Shock-tube results for case 2 at $t = 0.035$.(a) represents density, (b) represents velocity, (c) represents pressure, (d) represents energy.

In Figure 6 (Case 2), which corresponds to a flow regime characterized by a relatively large pressure gradient, the quantum solver maintains a high level of accuracy in resolv-ing the primary waveforms. However, as the solution approaches regions with steep spatial gradients, particularly at the leading and trailing edges of sharp fronts, small amplitude oscillations become visible. These oscillations are localized and do not significantly distort the bulk solution, but they indicate the onset of dispersion-like effects associated with the solver's handling of rapid transitions.

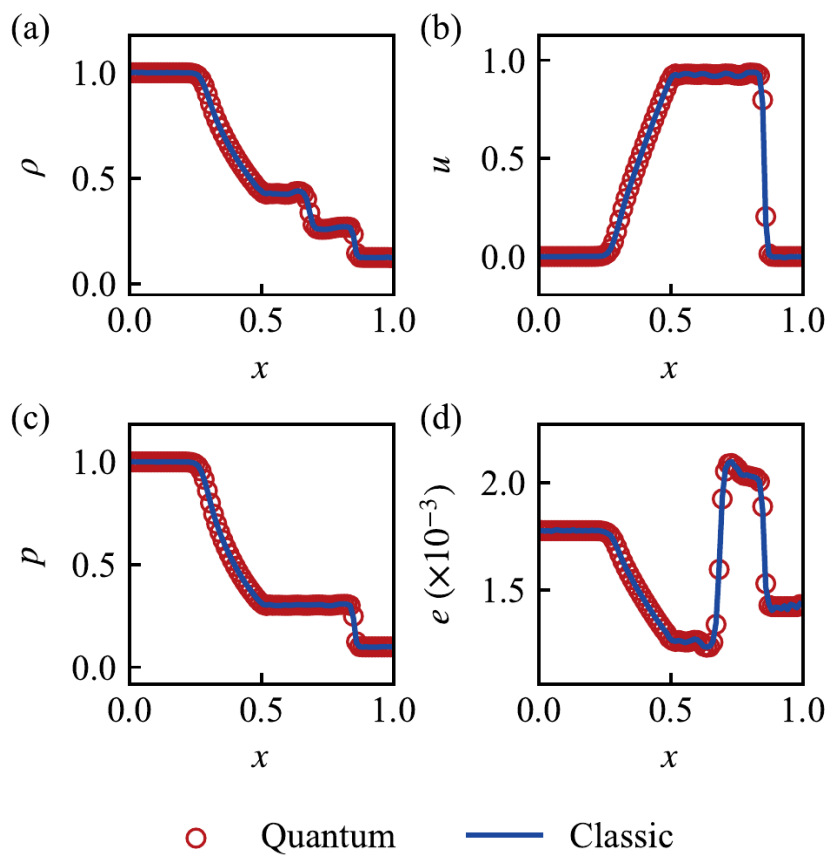

**Figure 5**  Shock-tube results for case 1 at $t = 0.2$. (a) represents density, (b) represents velocity, (c) represents pressure, (d) represents energy.

In Figure 7 (Case 3), representing the most extreme shock-dominated regime in this study, the solver continues to capture the correct large-scale shock positions, amplitudes, and overall wave topology. Nevertheless, the solution exhibits pronounced high-frequency artifacts superimposed on the shock profiles. These artifacts are more intense than in Case 2 and can be attributed to two primary factors: (i) the stochastic fluctuations introduced by quantum noise during measurement and state preparation, and (ii) limitations in the expressivity of the chosen ansatz, which constrain its ability to represent highly discontinuous features without introducing spurious modes. The combination of these effects leads to a slight degradation in local fidelity while preserving the global shock structure.

manifest as localized oscillations, particularly near discontinuities.

To assess temporal stability, figure 8 presents the time evolution of the shock-tube solution for case 1. Figures 8 (a),(c),(e) present the VQLS results as described in the text, while the remaining figures show classical reference solutions. As nonlinear features of the solution become increasingly pronounced, the VQLS results exhibit deviations in regions of high physical gradients due to its inherent error propagation and accumulation mechanism, while maintaining excellent agreement with classical solutions in other regions.

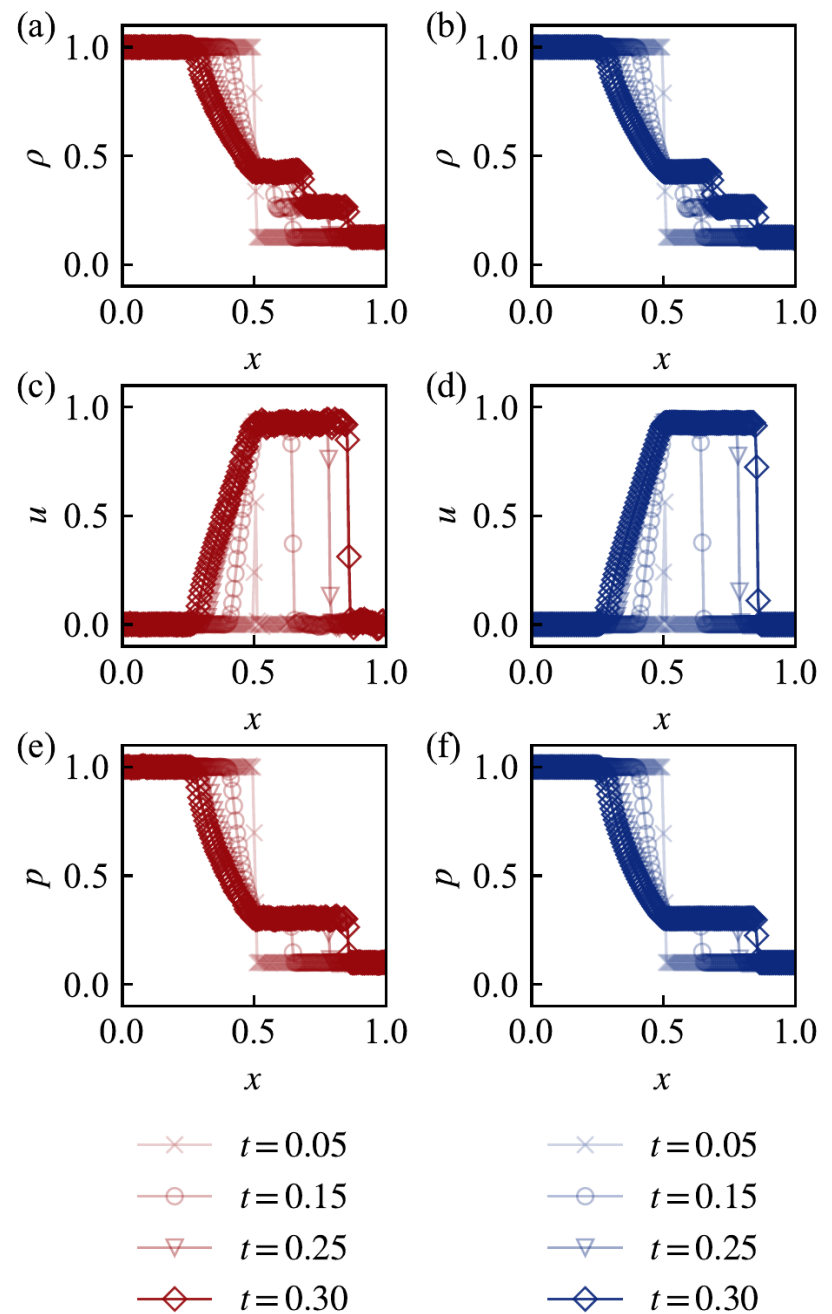

**Figure 8** Temporal evolution of the density field for case 1: comparison between quantum and classical results.

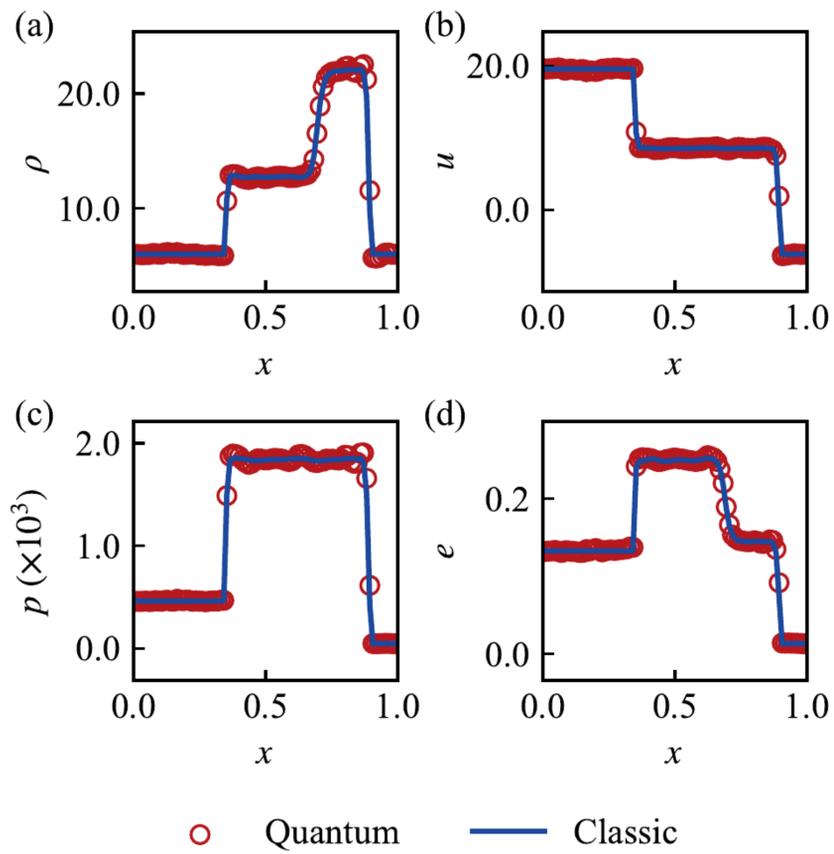

**Figure 7** Shock-tube results for case 3 at $t = 0.035$.(a) represents density, (b) represents velocity, (c) represents pressure, (d) represents energy.

Across all cases, the quantum solver demonstrates reliable convergence in moderate flow regimes. Quantitative agreement with classical benchmarks is observed for cases 1 and 2. In case 3, which features sharp gradients and strong nonlinearities, discrepancies emerge due to VQLS residual errors and ansatz constraints. These deviations grow over time and

### 3.3 Convergence analysis under multiple ansatz trees and domain partitions

This section presents a systematic analysis of how ansatz tree multiplicity and spatial partitioning affect the convergence behaviour of VQLS applied to the NS equations. The results indicate that convergence becomes increasingly difficult as the system size grows, a trend associated with the rapid expansion of the underlying optimisation landscape. In particular, large-dimensional parameter spaces are susceptible to regions with vanishing gradients, often referred to as barren plateaux, which hinder optimisation progress.

To address this, multiple ansatz trees are introduced to enhance the expressivity of the variational circuit. Each tree encodes a different region of the Hilbert space, allowing for a broader and more flexible approximation of the solution space. This strategy improves convergence by increasing the likelihood of locating parameter regions with favourable gradient properties and by enabling multiple search trajectories. In the numerical experiments, linear systems of increasing size were generated by refining the CFD mesh, corresponding to circuits with 9 and 10 qubits. The VQLS was applied using various numbers of ansatz trees. The results, shown in figure 9, demonstrate that convergence is accelerated as the number of trees increases.

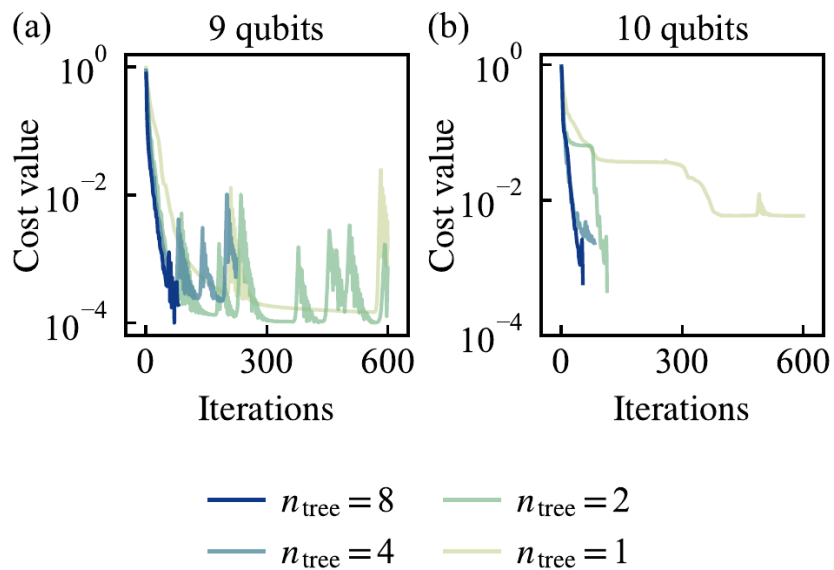

**Figure 9** Convergence of VQLS for different numbers of Ansatz trees. Subfigures (a) and (b) respectively display the convergence residual curves under different numbers of Ansatz trees, with the comparison between (a) and (b) demonstrating the computational efficiency using varying numbers of qubits.

In addition to circuit structure, spatial partitioning also influences convergence. By dividing the computational domain into smaller blocks, the dimensionality of each subproblem is reduced, simplifying the associated quantum circuit. These subdomains can be optimised independently and later combined to construct a global solution. This approach not only improves convergence but also allows better scalability on quantum hardware. Figure 10 shows the convergence histories for the system at $T = 10$ (Case 1), under different con-figurations of ansatz tree numbers and block counts. The results indicate a consistent trend: increasing both the number of trees and the number of blocks improves convergence.

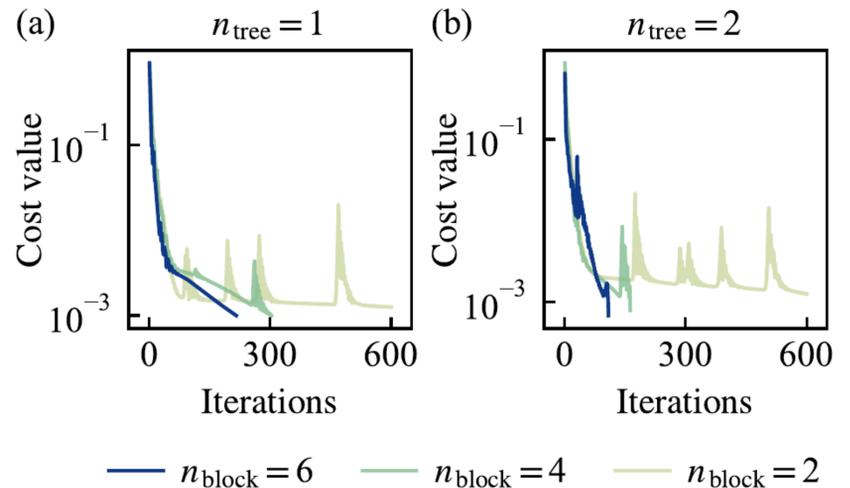

**Figure 10** Effect of ansatz tree number and partitioning on convergence. Increasing both the number of trees and the number of blocks improves convergence.

For reference, the classical LU decomposition used in this work factorises the matrix $A$ into lower and upper triangular matrices via Gaussian elimination, with overall complexity $O(N^3)$. In contrast, the VQLS without QRAM has an asymptotic complexity of $O(N + \kappa^2 \epsilon^{-2} \delta^{-1} \operatorname{poly}(\log N))$, where $\kappa$ is the condition number, $\epsilon$ the sampling precision, and $\delta$ the convergence threshold. Provided that a suitable classical optimiser and preconditioning strategy are applied, the quantum approach may offer advantages for sufficiently large $N$.

In summary, the combined use of multiple ansatz trees and domain partitioning improves the convergence and scalability of quantum solvers for fluid dynamics applications. The findings suggest that these strategies are promising for extending quantum methods to high-dimensional problems encountered in compressible flow simulations.

## 4. Conclusions

This work develops a VQLS framework incorporating a multi-ansatz tree architecture for the simulation of compressible flows governed by the one-dimensional NS equations. The method is tailored to the algebraic systems arising from implicit time integration schemes and is implemented within a hybrid quantum–classical computational loop. By combining multiple parameterised quantum circuits with classically optimised weighting coefficients, the framework expands the representational capacity of variational solvers without increasing circuit depth, thereby enhancing compatibility with current NISQ devices.

The proposed method is validated through canonical shock tube simulations under three distinct flow configurations, including high-Mach-number and viscous cases. Results demonstrate that the multi-ansatz approach consistently recovers the primary shock, rarefaction, and contact discontinuities with high accuracy when compared against classical solvers. Quantitative analyses reveal that the multi-ansatz architecture improves convergence rate and reduces residual errors relative to single-ansatz implementations. In particular, the combination of ansatz diversification and domain decomposition yields a systematic mitigation of barren plateau effects, enabling convergence in scenarios where baseline VQAs fail to produce viable solutions within a fixed iteration budget.

Furthermore, the algorithm exhibits favourable scaling properties with respect to both quantum circuit size and input problem dimensionality. Empirical observations indicate that increasing the number of ansatz branches leads to monotonic reductions in optimisation error up to a threshold determined by hardware noise and qubit count. The subdomain partitioning strategy further contributes to convergence acceleration by lowering the effective dimensionality of each variational task, thus alleviating gradient vanishing and improving solution stability across time steps.

Taken together, these results suggest that the multi-ansatz variational framework provides a viable path toward realising quantum-assisted solvers for nonlinear fluid systems. The method balances computational efficiency with architectural tractability and offers a scalable foundation for extending quantum algorithms to higher-dimensional and time-dependent NS problems. Ongoing work will focus on implementing the framework on real quantum hardware, exploring adaptive ansatz selection, and integrating quantum error mitigation strategies to further enhance solution fidelity under realistic noise conditions.

**Conflict of interest** On behalf of all authors, the corresponding author states that there is no conflict of interest.

**Data availability** The data that support the findings of this study are available from the corresponding author upon reasonable request.

**Author contributions** The research was conceptualised and designed by S. Yao and W. Zhao. The numerical framework was developed, implemented, and validated by S. Yao and Z. Duan, who also performed the simulations, processed the data, and prepared the initial manuscript draft. Z. Wang contributed to the structural organisation and refinement of the manuscript. W. Zhao and S. Xiong provided overall supervision, secured project funding, and led the critical revision and final editing of the manuscript.

**Supporting information** This research was supported by the National Key Research and Development Program of China (grant no. 2023YFB4502600), National Natural Science Foundation of China (grant nos. 92271114, 92371206, 92271204), and the Research Fund of the National Key Laboratory of Aerospace Physics in Fluids (grant no. 2024-APF-KFZD-07). The authors also thank Professor Chao Song and PhD candidate Jianan Yang (School of Physics, Zhejiang University) for their valuable discussions.